# Emergence of nanoscale inhomogeneity in the superconducting state of a homogeneously disordered conventional superconductor, NbN

Anand Kamlapure, Tanmay Das[*], Somesh Chandra Ganguli, Jayesh B Parmar, Somnath Bhattacharyya[†] and Pratap Raychaudhuri[‡]

*Tata Institute of Fundamental Research, Homi Bhabha Road, Colaba, Mumbai 400005, India.*

The notion of spontaneous formation of an inhomogeneous superconducting state is at the heart of most theories attempting to understand the superconducting state in the presence of strong disorder. Using scanning tunneling spectroscopy and high resolution scanning transmission electron microscopy, we experimentally demonstrate that under the competing effects of strong homogeneous disorder and superconducting correlations, the superconducting state of a conventional superconductor, NbN, spontaneously segregates into domains. Tracking these domains as a function of temperature we observe that the superconducting domains persist across the bulk superconducting transition, $T_c$, and disappear close to the pseudogap temperature, $T^*$, where signatures of superconducting correlations disappear from the tunneling spectrum and the superfluid response of the system.

[*] e-mail: tanmay@tifr.res.in
[†] e-mail: somnath@tifr.res.in
[‡] e-mail: pratap@tifr.res.in

The emergence of an inhomogeneous superconducting state in the presence of strong disorder is a recurring theme in the study of strongly disordered superconductors. When disorder is homogeneously distributed over atomic length scales, one would expect that its effect on the superconducting state will get averaged over length scales of the order of the superconducting coherence length (ξ), leaving a spatially uniform superconducting ground state. Quite early on, it was conjectured that for such a state the pairing energy scale, and consequently the superconducting transition temperature, $T_c$, will remain finite even in the limit of very strong disorder[1,2]. However, more recent numerical simulations[3,4,5] indicate that even in the presence of homogeneous disorder the superconducting state can segregate into superconducting and non-superconducting regions both made of Cooper pairs, where global phase coherence is achieved below $T_c$ through Josephson coupling between the superconducting domains. From a theoretical standpoint it is now understood that although the pairing energy scale can remain finite even at strong disorder[6,7], phase fluctuations between the superconducting domains can drive the transition from superconducting to non-superconducting state. Indeed, recent observation of a pseudogap state in several strongly disordered s-wave superconductors[8,9,10] (NbN, TiN and InO$_x$), characterized by a soft gap in the electronic energy spectrum which persists up to a temperature $T^*$, well above $T_c$, is consistent with this scenario.

While the formation of an inhomogeneous superconducting state has been invoked to explain a variety of phenomena close to the critical disorder for the destruction of the superconducting state e.g. magnetic field tuned superconductor-insulator transition[11,12], finite superfluid stiffness[13] above $T_c$ and Little-Parks oscillations in a disorder driven insulating film[14,15], a direct experimental proof of emergent inhomogeneity in a system where the structural disorder is indeed homogeneously distributed over typical length scale of superconducting domains is currently lacking. In this article we use a combination of high resolution scanning transmission electron microscopy (HRSTEM) and scanning tunneling spectroscopy (STS) to carry out a detailed study of the structural disorder vis-à-vis the inhomogeneity observed in the superconducting state in strongly disordered NbN thin films. The central result of this paper is that while structural disorder in these films is uniformly distributed over atomic length scales, the superconducting state segregates into domains, tens of nanometer in size. These domains continue to persist across $T_c$, and gradually become diffuse and disappear only close to $T^*$. These results provide a direct evidence of the persistence of superconducting puddles in the pseudogap state and a real space perspective of the nature of the superconducting phase transition in a strongly disordered s-wave superconductor.

**Results**

**Epitaxial NbN films as a homogeneously disordered superconductor.** Samples used in this study consist of 50 nm thick epitaxial NbN films. The films were grown using reactive magnetron sputtering on (100) oriented MgO single crystalline substrates by sputtering a high-purity niobium target in Ar/N$_2$ gas mixture[16]. Disorder in this system is in the form of Nb vacancies in the crystalline NbN lattice which is tuned by controlling the sputtering power and the Ar/N$_2$ ratio, which in turn determines the Nb/N ratio in the plasma. The effective disorder in each film was characterized[17] through the product of the Fermi wave vector ($k_F$) and electronic mean free path (*l*). At optimal deposition conditions, clean NbN film has $T_c \sim 16K$ corresponding to $k_Fl \sim 10$. As disorder is introduced in the system $T_c$ decreases monotonically down to the limit where superconductivity is completely destroyed down to 300 mK [Ref. 17]. In this study we concentrate on three samples with $T_c \sim$ 1.65 K, 2.9 K and 3.5 K with corresponding $k_Fl \sim$ 1.5-2.

The sample thickness is much larger than the corresponding dirty limit coherence length[18], $\xi \sim 6$ nm.

**Investigation of structural disorder at atomic length scales.** To characterize the disorder of the sample at the atomic level HRSTEM measurements were performed on a strongly disordered sample ($T_c \sim 2.5$ K, $k_F l \sim 1.7$) deposited under conditions identical to the one used for STS measurements. For reference, identical measurements were also performed on a low disordered sample with $T_c \sim 16$ K. To make the interfaces 'edge on', i.e, perpendicular to the incoming electron beam, both the samples were tilted along <110> in this present study. The structure of NbN projected along <110> is shown in the inset of Fig. 1a which reveals that atomic columns contain either Nb or, N i.e, atomic columns with mixed atoms are not present while viewing through this direction. High resolution Z-contrast images of MgO/NbN thin film interfaces of both samples (Fig. 1a and 1b) show that the films grow epitaxially on MgO (100) substrate. At low magnification (Fig. 1(c) and 1(d) ) the two samples look similar: After a distance of columnar kind growth the films grow uniformly.

The essential difference between the two samples is brought out when disorder is investigated at atomic length scales. For this purpose HRSTEM images were acquired at several locations for each sample in the uniform regions of the films shown in Fig 1c and 1d. HRSTEM images at two locations on each sample along with the corresponding surface plots of two dimensional intensity distributions are shown in Fig. 2 and Fig. 3. For all experiments, small camera length was purposefully selected, which allowed the high angle annular dark field (HAADF) detector to collect mainly electrons scattered at high angles which are mostly contributed by atomic columns containing Nb ($Z=41$). In this case, the intensity ($I$) of an atomic column in HRSTEM image is proportional to the number of Nb atoms ($n$) in the column[19,20].

Therefore, the intensity variation in these images reflects the variation of number of Nb atoms in respective columns resulting from Nb vacancies in the crystalline lattice. The smooth intensity variation in the low disordered sample (Fig. 2b and 2d) is primarily due to the overall thickness variation of TEM sample produced during ion milling. In contrast, in the strongly disordered sample (Fig. 3b and 3d) we observe large intensity variation even in adjacent columns, showing that Nb vacancies are randomly distributed in the crystalline lattice. Thus even in strongly disordered NbN thin films, structural disorder stems from randomly distributed Nb vacancies, while the films remains homogeneous when averaged over length scales larger than few nm.

**Coherence peaks as a measure of the superconducting order parameter (OP).** Direct access to the spatial inhomogeneity in the superconducting state is offered by STS measurements using a scanning tunneling microscope (STM). The tunneling conductance ($G(V)$) between the normal metal tip and the superconductor is given by[21],

$$G(V) \propto \frac{1}{R_N} \int_{-\infty}^{\infty} N_s(E) \left( -\frac{\partial f(E-eV)}{\partial E} \right) dE, \quad (1)$$

where $E$ is the energy measured from the Fermi energy, $N_s(E)$ is the single-particle local density of states (LDOS) in the superconductor and $f(E)$ is the Fermi-Dirac distribution function. In the limit $T \to 0$, the tunneling conductance, $dI/dV$, measured as a function of the voltage ($V$) between a normal metal tip and the sample gives the local density states (LDOS) on the surface. The spatial variation of tunneling conductance thus reflects the variation of the LDOS in an inhomogeneous system.

We first concentrate on the nature of individual tunneling spectra. Figures 4(a) and 4(c) show two representative spectra recorded at 500 mK at two different locations on the sample with $T_c \sim 2.9K$. The spectra show two common features: (i) A dip associated with the superconducting energy gap and (ii) a broad, temperature independent V-shaped background extending up to high bias arising from Alshuler-Aronov type electron-electron interactions. However, they differ strongly in the height of the coherence peak. While the spectrum in Fig. 4(a) displays well defined coherence peaks at the gap edge, in Figure 4(c) the coherence peak is completely suppressed. This difference becomes more prominent once the superconducting feature is isolated from the V-shaped background, by dividing the spectra obtained at low temperature by the spatially averaged spectrum at 9K, $G_{9K}(V)$, (Fig. 4(b)-(d)) where superconducting correlation are completely suppressed. Unlike "clean" NbN films where the spectrum is fully gapped[17], at strong disorder all the normalized spectra show significant zero bias conductance (ZBC). This is a general feature observed in all strongly disordered NbN thin films that we have measured. Barring an overall slope which is an artifact of the normalizing procedure, both kinds of spectra show a soft gap below 1 meV. In the rest of the paper we will refer to the background corrected and slope compensated spectra as the "normalized spectra", $G_N(V)$.

The density of states of a conventional clean superconductor, well described by the Bardeen-Cooper-Schrieffer (BCS) theory, is characterized by an energy gap (Δ), corresponding to the pairing energy of the Cooper pairs and two sharp coherence peaks at the edge of the gap, associated with the long-range phase coherent superconducting state. This is quantitatively described by a single particle DOS of the form[22],

$$N_s(E) = Re\left( \frac{|E| + i\Gamma}{\sqrt{(|E| + i\Gamma)^2 - \Delta^2}} \right), \quad (2)$$

where the additional broadening parameter Γ phenomenologically takes into account broadening of the DOS due recombination of electron and hole-like quasiparticles. For Cooper pairs without phase coherence, it is theoretically expected that the coherence peaks will get suppressed whereas the gap will persist[7]. Therefore, we associate the two kinds of spectra with regions with coherent and incoherent Cooper pairs respectively[10]. The normalized tunneling spectra with well defined coherence peaks can be fitted well within the BCS-Γ formalism using eq. 1 and 2. Fig. 5(a), 5(c) and 5(e) show the representative fits for the three different samples. In all the samples we observe Δ to be dispersed between 0.8-1.0 meV corresponding to a mean value of $2\Delta/k_BT_c$ ~ 12.7, 7.2 and 6 (for $T_c$ ~ 1.65 K, 2.9 K and 3.5 K respectively) which is much larger than the value 3.52 expected from BCS theory[21]. Since Δ is associated with the pairing energy scale, the abnormally large value of $2\Delta/k_BT_c$ and the insensitivity of Δ on $T_c$ suggest that in the presence of strong disorder $T_c$ is not determined by Δ. On the other hand, Δ seems to be related to $T^*$ ~ 7-8 K which gives $2\Delta/k_BT^*$ ~ 3.0±0.2, closer to the BCS estimate. Γ/Δ is relatively large and shows a distinct increasing trend with increase in disorder. In contrast, spectra that do not display coherence peaks (Fig. 5(b), 5(d) and 5(f)) cannot be fitted using BCS-Γ form for DOS. However, we note that the onset of the soft-gap in this kind of spectra happens at energies similar to the position of the coherence peaks, showing that the pairing energy is not significantly different between points with and without coherence.

Since the coherence peaks are directly associated with phase coherence of the Cooper pairs, the height of the coherence peaks provides a direct measure of the local superconducting order parameter. This is consistent with numerical Monte Carlo simulations[7] of disordered superconductors using attractive Hubbard model with random on-site disorder which show that the coherence peak height in the LDOS is directly related to the local superconducting OP $\Delta_{OP}(R) = \langle c_{R\downarrow} c_{R\uparrow} \rangle$. Consequently, we take the average of the coherence peak height (*h*) at positive and negative bias (with respect to the high bias background) as an experimental measure of the local superconducting OP (Fig. 4(b)).

**Emergent inhomogeneity in the superconducting state.** To explore the emergent inhomogeneity we plot in Fig. 6(a), 6(b) and 6(c) the spatial distribution of *h*, measured at 500 mK in the form of intensity plots for three samples over 200 × 200 nm area. The plot shows large variation in *h* forming regions where the OP is finite (yellow-red) dispersed in a matrix where the OP is very small or completely suppressed (blue). The yellow-red regions form irregular shaped domains dispersed in the blue regions. The fraction of the blue regions progressively increases as disorder is increased. To analyze the spatial correlations we calculate the autocorrelation function (ACF), defined as, $\rho(\bar{x}) = \frac{1}{n(\sigma_h)^2} \sum_{\bar{y}} (h(\bar{y}) - \langle h \rangle)(h(\bar{y} - \bar{x}) - \langle h \rangle)$, where *n* in the total number of pixels and $\sigma_h$ is the standard deviation in *h*. The circular average of $\rho(\bar{x})$ is plotted as a function of $|\bar{x}|$ in Fig. 6(j) showing that the correlation length becomes longer as disorder is increased. The domain size progressively decreases with decrease in disorder and eventually disappears in the noise level for samples with $T_c \gtrsim 6K$. From the length at which the ACF drops to the levels of the base line we estimate the domains sizes to be 50 nm, 30 nm and 20 nm for the samples with $T_c$ ~ 1.65 K, 2.9 K and 3.5 K respectively. The emergent nature of the superconducting domains is apparent when we compare structural inhomogeneity with the *h*-maps. While the defects resulting from Nb vacancies are homogeneously distributed over atomic length scales, the domains formed by superconducting correlations over this disordered landscape is 2 orders of magnitude larger.

The domain patterns observed in *h*-maps is also visible in Fig. 6(d), 6(e) and 6(f) when we plot the maps of zero bias conductance (ZBC), $G_N(0)$, for the same samples. The ZBC maps show an inverse correlation with the *h*-maps: Regions where the superconducting OP is large have a smaller ZBC than places where the OP is suppressed. The cross-correlation between the *h*-map and ZBC map can be computed through the cross-correlator defined as,

$$I = \frac{1}{n}\sum_{i,j}\frac{(h(i,j)-\langle h\rangle)(ZBC(i,j)-\langle ZBC\rangle)}{\sigma_h \sigma_{ZBC}},$$ where *n* is the total number of pixels and $\sigma_{ZBC}$ is the standard deviations in the values of ZBC. A perfect correlation (anti-correlation) between the two images would correspond to $I = 1(-1)$. We obtain a cross correlation, $I = -0.3$ showing that the anti-correlation is weak. Thus ZBC is possibly not governed by the local OP alone. This is also apparent in the 2-dimensional histograms of *h* and ZBC (Fig. 6(g), 6(h) and 6(i)) which show a large scatter over a negative slope.

The weak anti-correlation between *h* and ZBC allows us to track the domain structure at elevated temperature where the coherence peaks get diffused due to thermal broadening and the *h*-maps can no longer be used as a reliable measure of the OP distribution. We investigated the temperature evolution of the domains as a function of temperature for the sample with $T_c \sim 2.9$ K. The bulk pseudogap temperature was first determined for this sample by measuring the tunneling spectra at 64 points along a 200 nm line at ten different temperatures. Fig.7(a) shows the temperature evolution of the normalized tunneling spectra along with temperature variation of resistance. In principle, at the $T^*$, $G_N(V=0) \approx G_N(V>>\Delta/e)$. Since this cross-over point is difficult to uniquely determine within the noise levels of our measurements, we use $G_N(V=0)/G_N(V=3.5\ \mathrm{mV}) \sim 0.95$ as a working definition for the $T^*$. Using this definition we obtain $T^* \sim 7.2$ K for this sample.

Spectroscopic maps were subsequently obtained at 6 different temperatures over the same area as the one in Fig. 6(e). Before acquiring the spectroscopic map we corrected for the small drift using the topographic image, such that the maps were taken over the same area at every temperature. Fig. 7(b)-(g) show the ZBC maps as a function of temperature. Below $T_c$, the domain pattern does not show a significant change and for all points $G_N(V=0)/G_N(V=3.5\ mV) < 1$ showing that a soft gap is present everywhere. As the sample is heated across $T_c$ Most of these domains continue to survive at 3.6K across the superconducting transition. Barring few isolated points ($<5\%$) the soft gap in the spectrum persist even at this temperature. At 6.9K, which is very close to $T^*$, most of the domains have merged in the noise background, but the remnant of few domains, originally associated with a region with high OP is still visible. Thus the inhomogeneous superconducting state observed at low temperature disappears at $T^*$.

**Discussion**

We now discuss the implication of our results on the nature of the superconducting transition. In a clean conventional superconductor the superconducting transition, well described through BCS theory, is governed by a single energy scale, $\Delta$, which represent the pairing energy of the Cooper pairs. Consequently, $T_c$ is given by the temperature where $\Delta \rightarrow 0$. This is indeed the case for NbN thin films in the clean limit. On the other hand in the strong disorder limit, the persistence of the gap in the single particle energy spectrum in the pseudogap state and the insensitivity of $\Delta$ on $T_c$ conclusively establishes that $\Delta$ is no longer the energy scale driving the superconducting transition. Indeed, the formation of an inhomogeneous superconducting state supports the notion that the superconducting state should be visualized as a disordered network of superconducting islands where global phase coherence is established below $T_c$ through Josephson tunneling between superconducting islands. Consequently at $T_c$, the phase coherence would get destroyed through thermal phase fluctuations between the superconducting domains, while coherent and incoherent Cooper pairs would continue to survive as evidenced from the persistence of the domain structure and the soft gap in the tunneling spectrum at temperatures above $T_c$. Finally, at $T^*$ we reach the energy scale set by the pairing energy $\Delta$ where the domain structure and the soft gap disappears.

These measurements connect naturally to direct measurements of the superfluid phase stiffness ($J$) performed through low frequency penetration depth and high frequency complex conductivity ( $\sigma = \sigma'(\omega) - i\sigma''(\omega)$ ) measurements on similar NbN samples. Low frequency measurements[8] reveal that in the same range of disorder where the pseudogap appears ( $T_c \lesssim$ 6K), $J(T \rightarrow 0)$ becomes a lower energy scale compared to $\Delta(0)$ (also see supplementary material). High frequency microwave measurements[13] reveal that in the pseudogap regime the superfluid stiffness becomes strongly frequency dependent. While at low frequencies $J$ ($\propto \omega\sigma''(\omega)$) becomes zero close to $T_c$ showing that the global phase coherent state is destroyed, at higher frequencies $J$ continues to remain finite up to a higher temperature, which coincides with $T^*$ in the limit of very high frequencies. Since the probing length scale set by the electron diffusion length at microwave frequencies[13] is of the same order as the size of the domains observed in STS, finite $J$ at these frequencies implies that the phase stiffness continues to remains finite within the individual phase coherent domains. Similar results were also obtained from the microwave complex conductivity of strongly disordered $InO_x$ thin films[23].

In summary, we have demonstrated the emergence of an inhomogeneous superconducting state, consisting of domains made of phase coherent and incoherent Cooper pairs in homogeneously disordered NbN thin films. The domains are observed both in the local variation of coherence peak heights as well as in the ZBC which show a weak inverse correlation with respect to each other. The origin of a finite ZBC at low temperatures as well as this inverse correlation is not understood at present and should form the basis for future theoretical investigations close to the SIT. However, the persistence of these domains above $T_c$ and subsequent disappearance only close to $T^*$ provide a real space perspective on the nature of the superconducting transition, which is expected to happen through thermal phase fluctuations between the phase coherent domains, even when the pairing interaction remains finite. However, an understanding of the explicit connection between this inhomogeneous state and percolative transport for the current above and below $T_c$ is currently incomplete[24,25,26], and its formulation would further enrich our understanding of the superconducting transition in strongly disordered superconductors.

**Methods**

**Scanning transmission electron microscopy measurements.** To prepare specimen for experiments using STEM both the samples were cut using Microsaw MS3 (Technoorg Linda, Hungary) with a dimension of 0.5 mm × 1.5 mm to make sandwich which was then placed within a 1mm × 1.8 mm slot of Titanium 3 slots grid (Technoorg Linda, Hungary) and fixed with G1 epoxy (Gatan Inc,USA). Subsequently grinding and dimpling were done to bring down the specimen thickness to a residual value of 10 to 15 µm, and finally Ar+ ion-beam thinning was performed. Substantial heating of the STEM foils and consequently the introduction of artifacts was avoided by means of double-sided ion-beam etching at small angles (< 6°) assisted by liquid $N_2$ cooling and by using low energies (acceleration voltage: 2.5 kV; beam current < 8 µA). A FEI-TITAN microscope operated at 300 kV equipped with FEG source, Cs (spherical aberration coefficient) corrector for condenser lens systems and a high angle annular dark field (HAADF) detector was used to perform STEM experiments. Sample was tilted to <110> zone axes to make the growth direction of the thin film structures parallel to the electron beam. Semi convergence angle of electron probe incident on the specimen and camera length were maintained at 17.8 mrad and 128 mm respectively during the experiments. All images were acquired with High Angle Annular Dark Field (HAADF) detector for about 20 microseconds.

**Scanning tunneling spectroscopy measurements.** STS measurements were performed on a home built scanning tunneling microscope (STM) operating down to 500 mK. The construction of the STM is similar to an earlier low temperature STM operating down to 2.6K [ref. 8] but coupled with a liquid $^3$He cryostat to achieve lower temperatures. For STS measurements NbN thin films were grown through reactive sputtering in an *in-situ* deposition chamber connected to the STM. The samples were deposited on single crystalline MgO substrates with pre-deposited contact pads mounted on a specially designed molybdenum sample holder[8] which allowed us to carry out the deposition at $600^0$ C. The temperature was accurately controlled through 2 resistive temperature sensors placed on the top and the bottom of the STM head and a resistive heater. We observe a small drift ~ 10 nm between 500 mK and 10 K. We correct this drift by tracking topographic feature before we start experiments at every temperature. For STS measurements, the differential conductance of the tunnel junction between the tip and the sample was measured using a low frequency *lock-in* technique. All STS maps presented here were acquired on a 32 × 32 grid over 200 × 200 nm area. For each sample the temperature variation of resistance was measured *ex-situ* after all STS measurements were completed.

---

[1] Anderson, P. W. Theory of Dirty Superconductors. J. Phys. Chem. Solids **11**, 26 (1959).

[2] Ma, M. and Lee, P. Localized superconductors. Phys. Rev. B **32**, 5658 (1985).

[14] Stewart Jr., M. D., Yin, A., Xu, J. M. and Valles Jr., J. M. Superconducting Pair Correlations in an Amorphous Insulating Nanohoneycomb Film, Science **318**, 1273 (2007).

[15] Kopnov, G. et al. Little-Parks Oscillations in an Insulator. Phys. Rev. Lett. **109**, 167002 (2012).

[16] Chockalingam, S. P. et al. Superconducting properties and Hall effect in epitaxial NbN thin films. Phys. Rev. B **77**, 214503 (2008).

[17] Chand, M. et al. Phase diagram of a strongly disordered s-wave superconductor, NbN, close to the metal-insulator transition. Phys. Rev. B **85**, 014508 (2012).

[18] Chand, M. *Ph.D. thesis,* Tata Institute of Fundamental Research, Mumbai, India. (available at http://www.tifr.res.in/~superconductivity/pdfs/madhavi.pdf )

[19] Wang, Z.L., Cowley, J.M. Simulating high-angle annular dark-field stem images including inelastic thermal diffuse scattering. Ultramicroscopy **31**, 437 (1989).

[20] LeBeau, J.M., Findlay, S.D., Allen, L.J., Stemmer, S. Standardless Atom Counting in Scanning Transmission Electron Microscopy. Nano Letters **10**, 4405 (2010).

[21] Tinkham, M *Introduction to Superconductivity* (Dover Publications Inc., Mineola, New York, 2004).

[22] Dynes, R. C., Narayanamurti, V., and Garno, J. P. Direct Measurement of Quasiparticle-Lifetime Broadening in a Strong-Coupled Superconductor. Phys. Rev. Lett. **41,** 1509 (1978).

[23] Liu, W., Kim, M., Sambandamurthy, G. and Armitage, N. P. Dynamical study of phase fluctuations and their critical slowing down in amorphous superconducting films. Phys. Rev. B **84**, 024511(2011).
15

**Acknowledgements**

We would like to thank Lara Benfatto, Nandini Trivedi, Karim Bouadim, Sudhansu Mandal and Vikram Tripathi for useful discussions, and Vivas Bagwe and John Jesudasan for helping with the experiments and Subash Pai from Excel Instruments, Mumbai for continuous technical support..

**Author contributions**

AK and TD contributed equally as first authors. PR and SB conceptualized the problem. JBP prepared the sample for TEM measurements. TD and SB performed the TEM measurements and the corresponding data analysis. AK designed and built the low temperature STM and AK and SCG performed the STS measurements and the corresponding data analysis under the supervision of PR. SB and PR co-wrote the paper. All authors discussed the results and commented on the analysis.

**Additional Information**

The authors declare no competing financial interest.

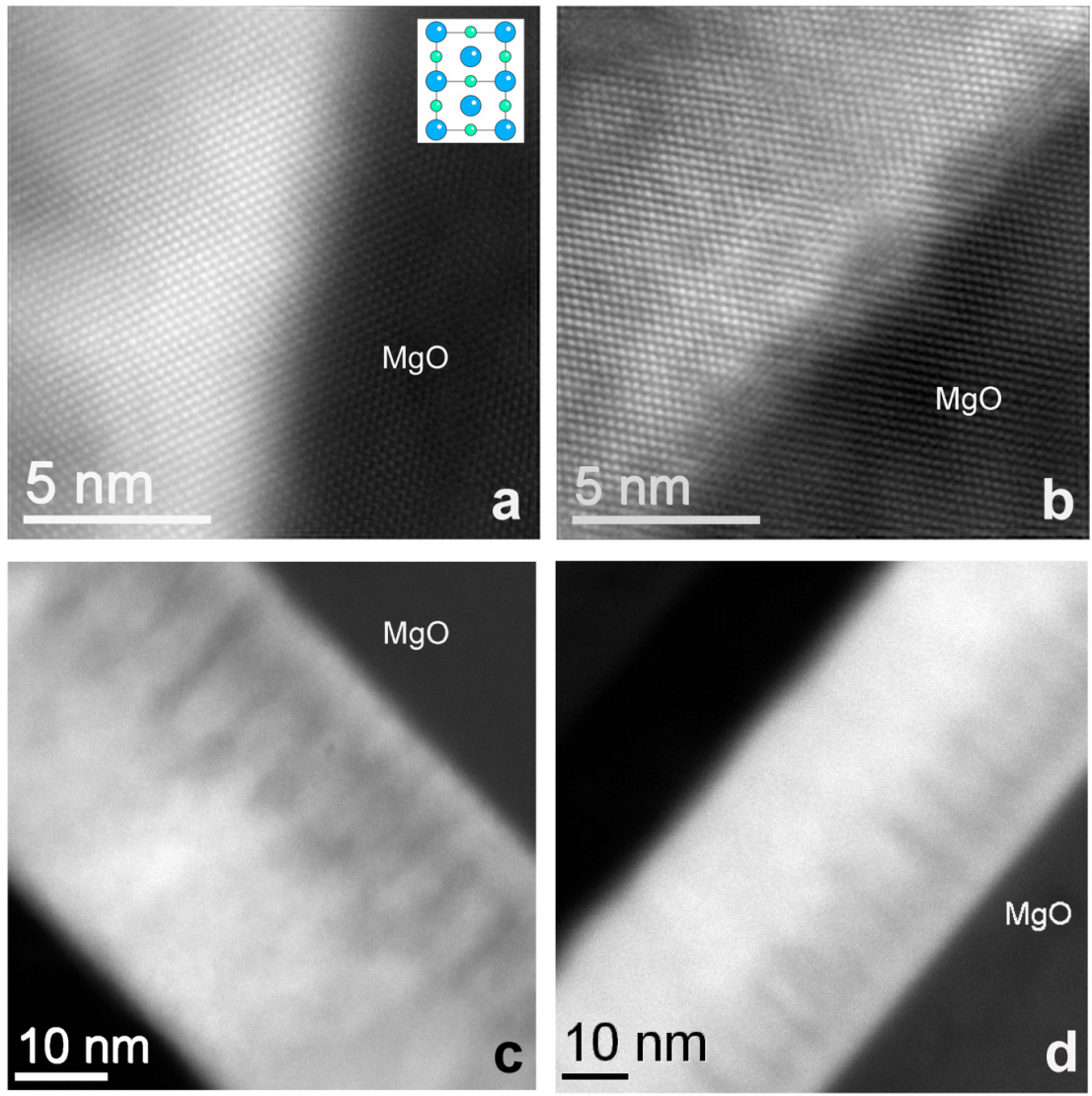

**Figure 1| Epitaxial growth of NbN thin films.** HRSTEM images showing epitaxial growth of NbN on MgO for samples having $T_c$ of (a) 16 K and (b) 2.5 K. Crystal structure of NbN projected along <110> is shown in inset of (a) where blue and green circles represent Nb and N atoms respectively. (c)-(d) Low magnification images for the same samples as in panel (a) and (b) respectively.

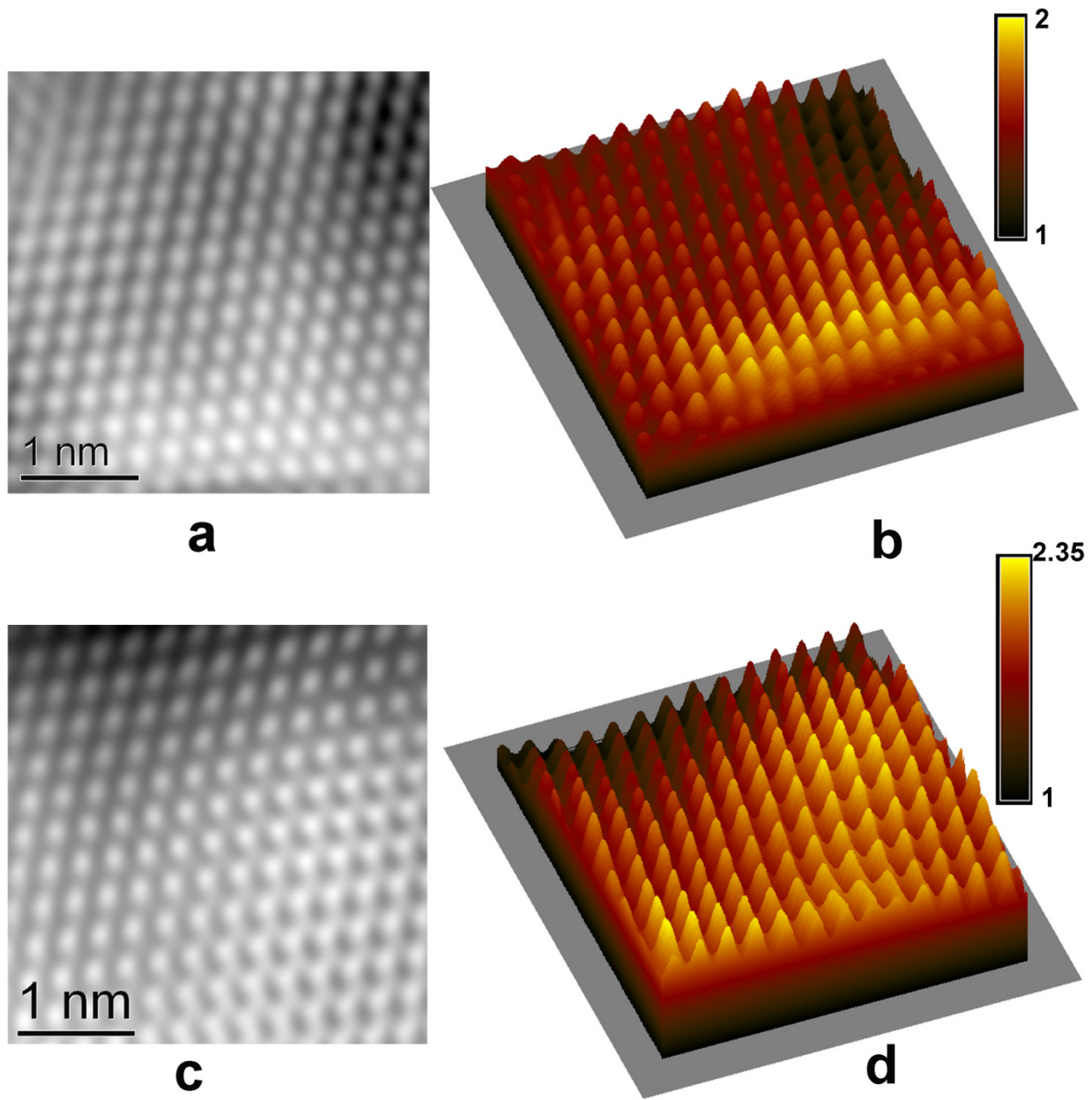

**Figure 2| Atomic scale structural characterization of NbN film with $T_c \sim 16K$.** (a), (c) HRSTEM images at two different locations for sample having $T_c \sim 16$ K and (b), (d) corresponding surface plots of the two dimensional intensity distributions. Each of the intensity distribution is normalized with respect to the minimum intensity value of the corresponding imaged region.

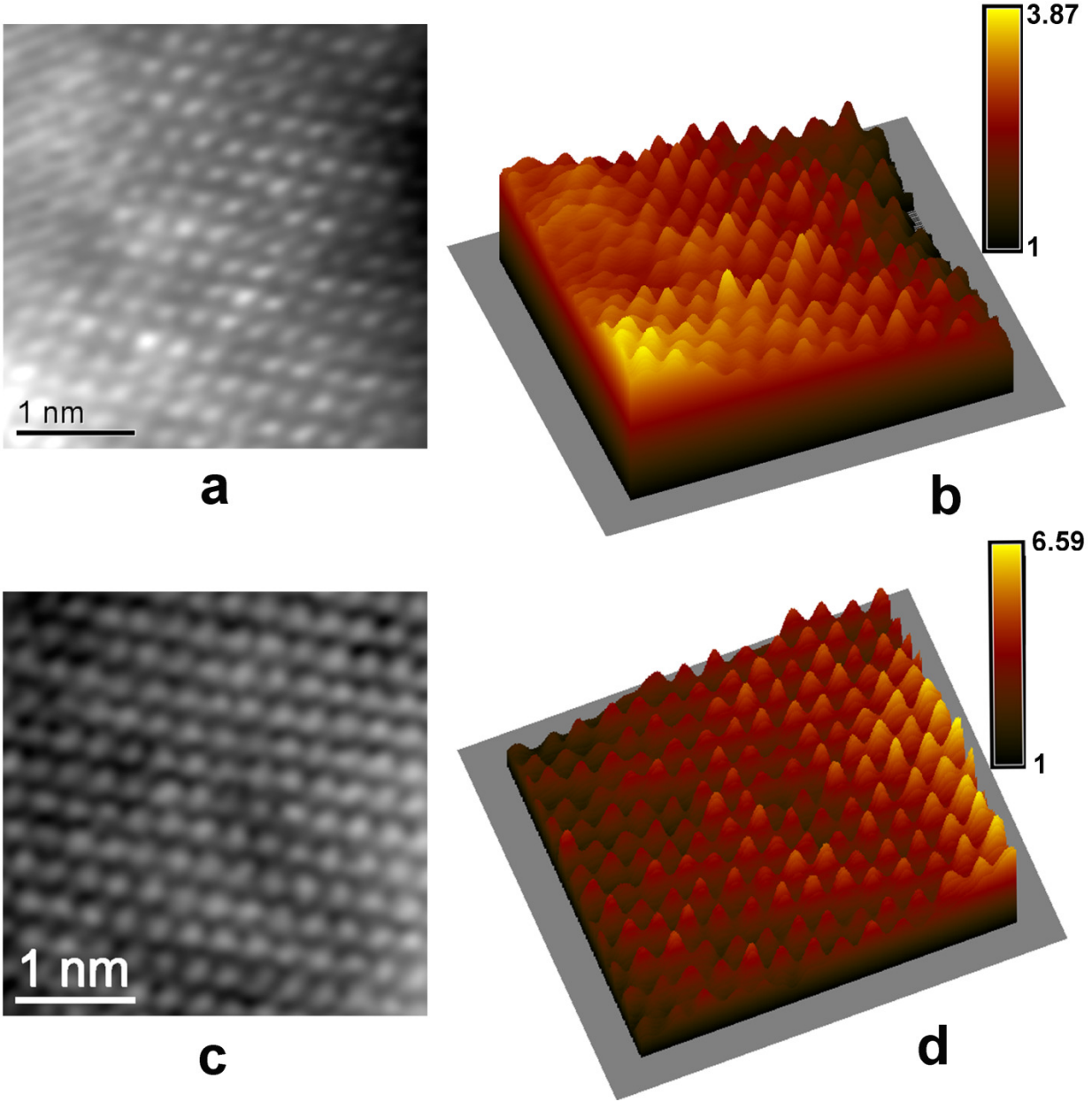

**Figure 3| Atomic scale structural characterization of NbN film with $T_c \sim 2.5$ K.** (a) & (c) HRSTEM images at two different locations for sample having $T_c \sim 2.5$ K and (b) & (d) corresponding surface plots of the two dimensional intensity distributions. Each of the intensity distribution is normalized with respect to the minimum intensity value of the corresponding imaged region.

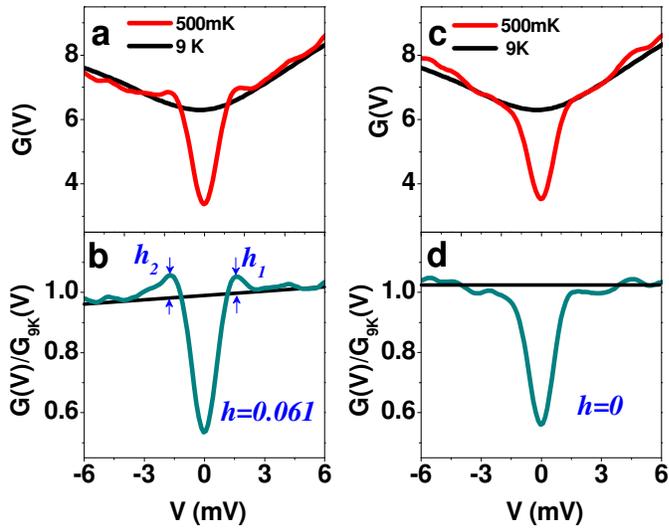

**Figure 4| Two kinds of tunneling spectra.** (a), (c) Red curves are the $G(V) = dI/dV$ vs. $V$ tunneling spectra acquired at 500 mK at two different locations on the sample with $T_c \sim 2.9$ K. Black curve represents the spatially averaged spectrum on the same sample acquired at 9K. (b), (d) Background corrected spectra corresponding to (a) and (c) respectively; the coherence peak height, $h$, is determined by measuring the average height of the peaks at positive ($h_1$) and negative ($h_2$) bias from a straight line (black) passing through high bias region.

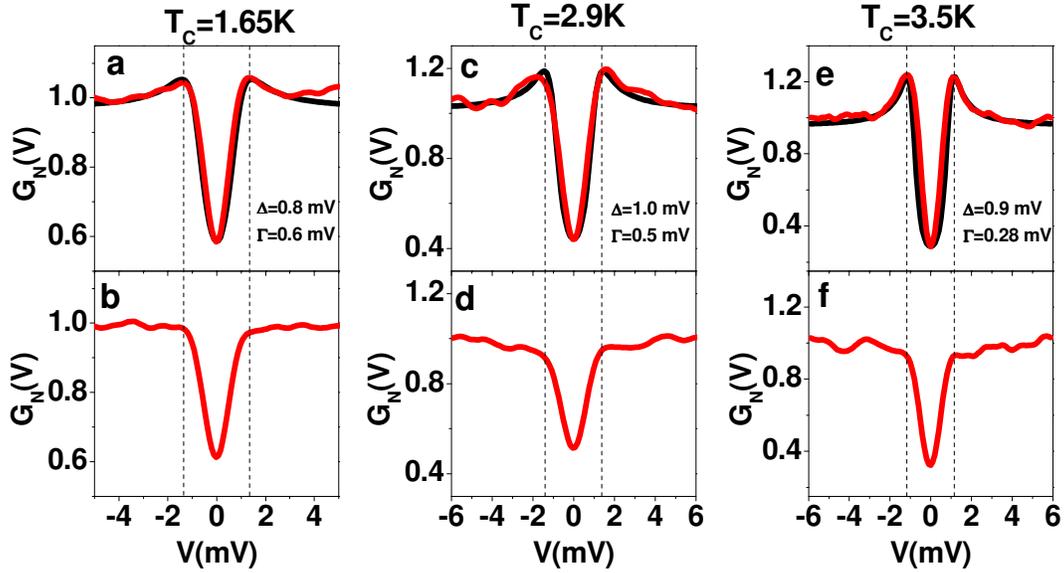

**Figure 5| Pairing energy and the onset of the soft gap in representative spectra for three samples with $T_C$=1.65K, 2.9K and 3.5K.** (a), (c), (e) Normalized tunneling spectra (red) on three different sample exhibiting well defined coherence peaks. Black curves correspond to the BCS-Γ fits using the parameters shown in each panels. (b), (d), (f) Normalized tunneling spectra at a different location on the same samples as shown in (a)-(c) showing no coherence peaks; note that the onset of the soft gap in these spectra coincide with the coherence peak positions in (a)-(c).

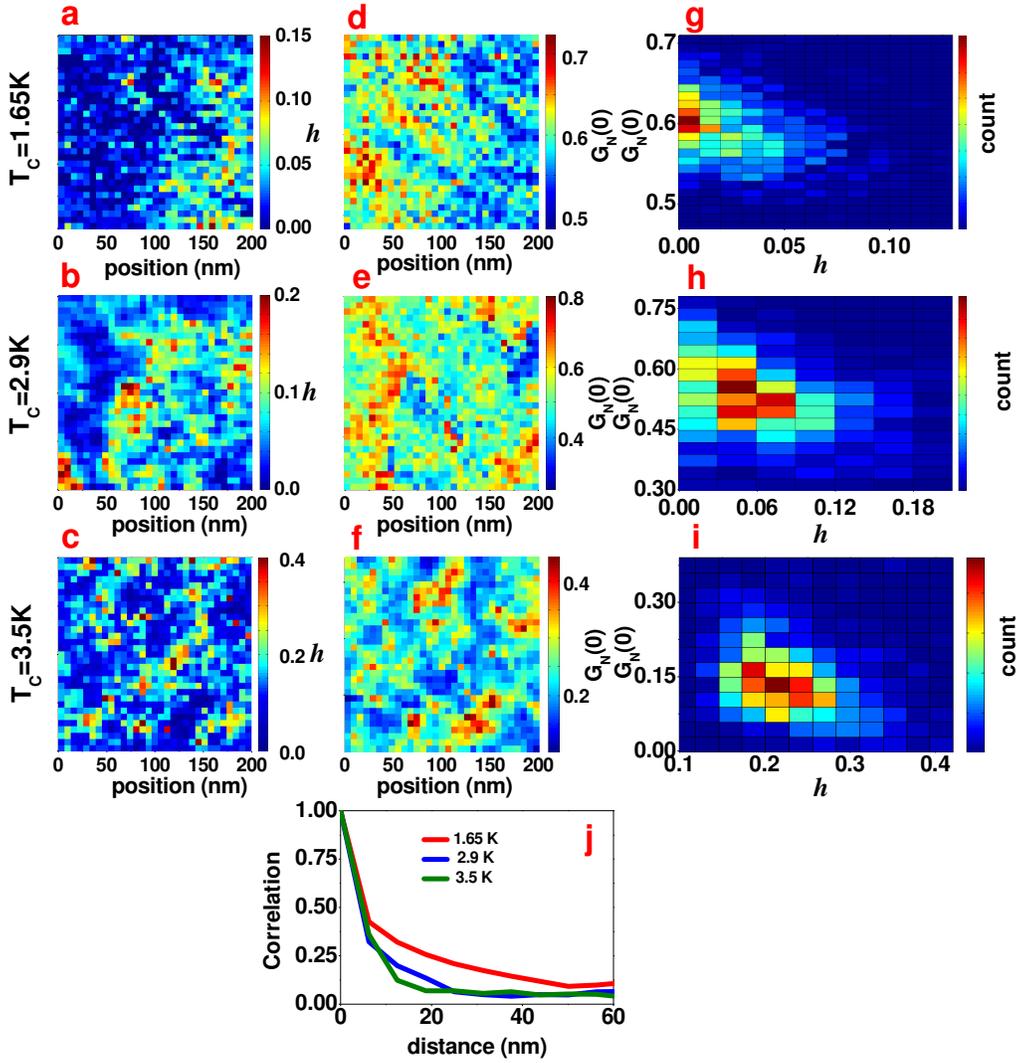

**Figure 6| Emergent inhomogeneity from STS at 500 mK in three samples with $T_c \sim$ 1.65K, 2.9K and 3.5K.** (a)-(c) Spatial variation of *h* in the form of intensity plot over 200 × 200 nm area. (d)-(f) spatial variation of ZBC ($G_N(V=0)$) over the same area for the same three samples. (g)-(i) 2-dimensional histogram of *h* and $G_N(V=0)$ showing weak anti-correlation between the two quantities. The values of $T_c$ corresponding to each row for panels (a)-(i) are given on the left side of the figure. (j) Radial average of the 2-dimensional autocorrelation function plotted as a function of distance for the three samples.

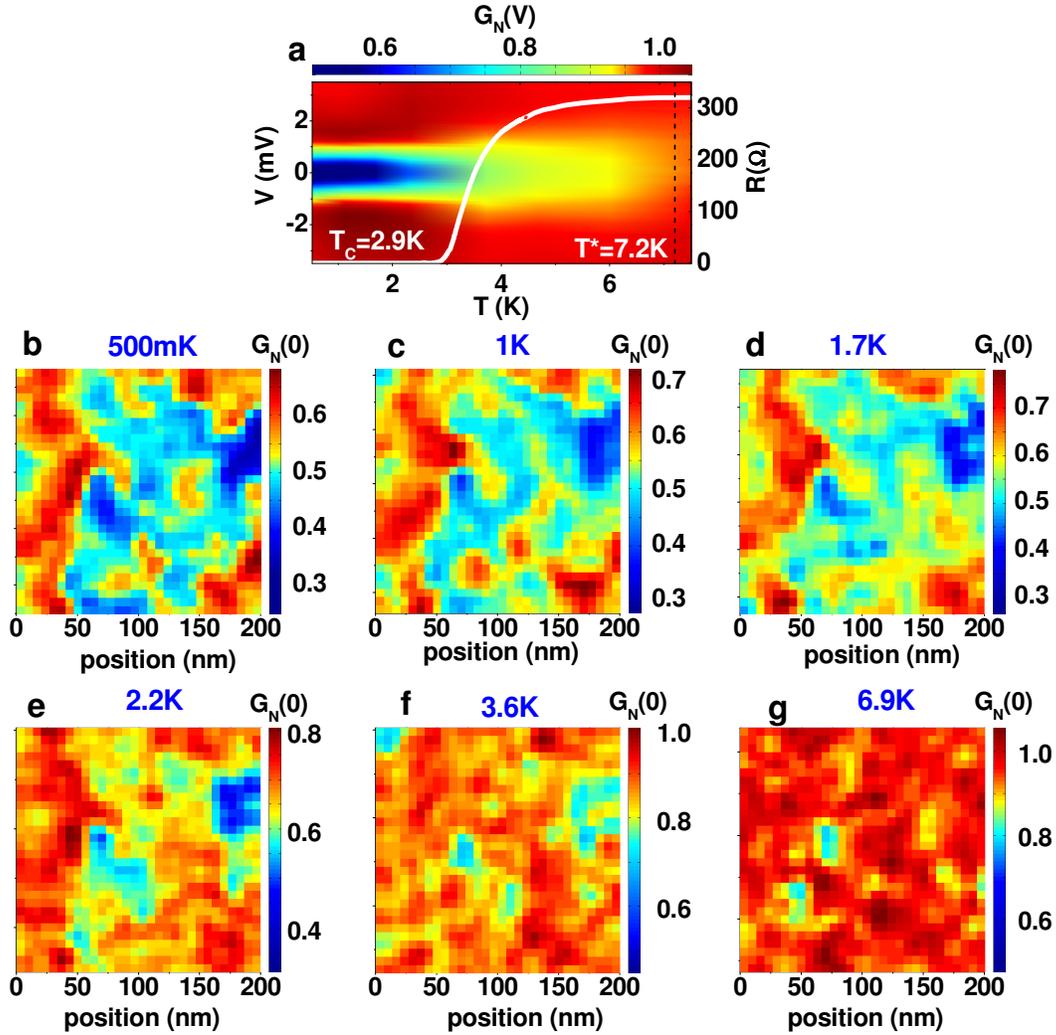

**Figure 7| Temperature evolution of the inhomogeneous superconducting state for the sample with $T_C$=2.9K.** (a) Temperature evolution of spatially averaged normalized tunneling spectra plotted in the form of intensity plot of $G_N(V)$ as a function of bias voltage and temperature. Resistance vs temperature (R-T) for the same sample is shown in white curve on the same plot. Pseudogap temperature $T^*$ ~ 7.2 K is marked with the dotted black line on top of the plot. (b)-(g) Spatial variation of ZBC ($G_N(V=0)$) plotted in the form of intensity plot over the same area for six different temperatures.

**Supplementary Material**

# Emergence of nanoscale inhomogeneity in the superconducting state of a homogeneously disordered conventional superconductor

Anand Kamlapure, Tanmay Das, Somesh Chandra Ganguli, Jayesh B Parmar, Somnath Bhattacharyya and Pratap Raychaudhuri

*Tata Institute of Fundamental Research, Homi Bhabha Road, Colaba, Mumbai 400005, India.*

**Energy and temperature scales as a function of disorder in NbN thin films**

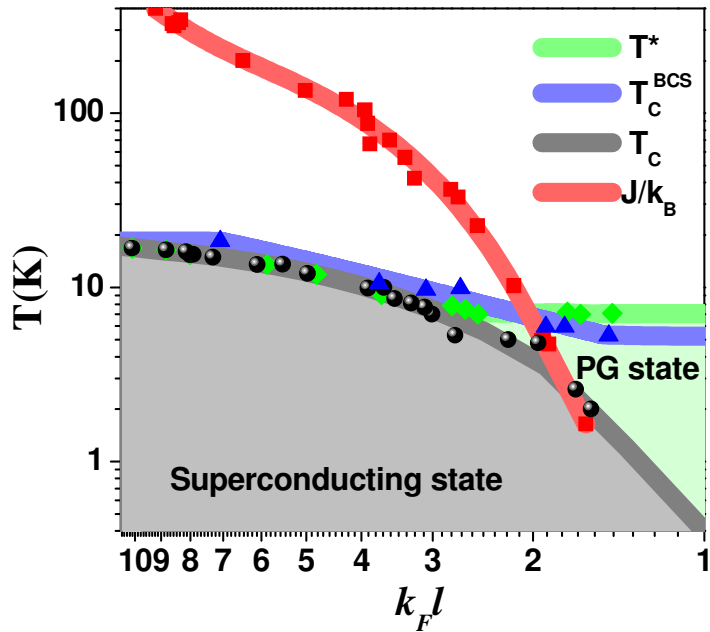

In the above phase diagram we summarize the evolution of various temperature and energy scales as a function of the effective disorder, $k_Fl$, in NbN thin films. $T_c$ (black circles), $T^*$ (green diamond) and the superfluid stiffness $J$ (red square), all in units of Kelvin, are taken from Ref. 1 and 2. $T_c^{BCS}$ is the BCS temperature scale associated with pairing energy, given by, $T_c^{BCS} = \Delta(0)/(1.76 k_B)$. Values of $\Delta(0)$ are taken from refs. 1, 3 (planar tunneling measurements), ref. 2 (STS measurements) and the present work. It is instructive to note in the range of disorder where the pseudogap appears, $T_c^{BCS}$ is closer to $T^*$ rather than $T_c$ as expected from BCS theory. In the same range of disorder $J$ become equal or smaller than $\Delta$ making the superconductor susceptible to phase fluctuations.

---

[1] Mondal, M. et al., Phase Fluctuations in a Strongly Disordered s-Wave NbN Superconductor Close to the Metal-Insulator Transition. Phys. Rev. Lett. **106**, 047001 (2011).

[2] Chand, M. et al. Phase diagram of a strongly disordered s-wave superconductor, NbN, close to the metal-insulator transition. Phys. Rev. B **85**, 014508 (2012).

[3] S. P. Chockalingam, S. P. et al. Tunneling studies in a homogeneously disordered s-wave superconductor: NbN. Phys. Rev. B **79,** 094509 (2009).